\documentclass[aps, prd, twocolumn,superscriptaddress,nofootinbib,10pt]{revtex4-2}
\usepackage{graphicx}
\usepackage{dcolumn}
\usepackage{multirow}
\usepackage[table]{xcolor}
\usepackage{colortbl,hhline}
\usepackage{amssymb,amsmath,amsthm,mathrsfs}
\usepackage{epstopdf}
\usepackage{hyperref}
\usepackage{empheq}
\usepackage{color}
\usepackage[normalem]{ulem} 
\usepackage{physics}
\usepackage[normalem]{ulem}
\usepackage[ruled,vlined]{algorithm2e}
\usepackage{enumitem}
\usepackage{empheq}

\usepackage{comment}
\usepackage{hyperref}
\usepackage[outercaption]{sidecap}
\newcommand{\be}{\begin{equation}}
\newcommand{\ee}{\end{equation}}
 \newcommand{\bea}{\begin{eqnarray}}
\newcommand{\eea}{\end{eqnarray}}

\usepackage{tikz}
\usetikzlibrary{calc,3d,arrows}
\usetikzlibrary{decorations.pathreplacing}
\usetikzlibrary{arrows.meta}

\newcommand{\RNum}[1]{\text{\uppercase\expandafter{\romannumeral #1\relax}}}

\definecolor{burntorange}{rgb}{0.8, 0.33, 0.0}

\begin{document}

\title{Gravitational Wave Emission from a Cosmic String Loop, I: Global Case}

\newcommand{\addressIFIC}{Instituto de F\'isica Corpuscular (IFIC), Consejo Superior de Investigaciones Cient\'ificas (CSIC) and Universitat de Val\`{e}ncia, 46980, Valencia, Spain}
\newcommand{\addressEHU}{Department of Physics, University of Basque Country, UPV/EHU, 48080, Bilbao, Spain}
\newcommand{\addressEHUQC}{EHU Quantum Center, University of Basque Country, UPV/EHU}
\newcommand{\addressNottingham}{School of Physics and Astronomy, University of Nottingham, Nottingham, NG7 2RD, UK}

\author{Jorge Baeza-Ballesteros}\email{jorge.baeza@ific.uv.es} \affiliation{\addressIFIC} 
\author{Edmund J. Copeland}\email{edmund.copeland@nottingham.ac.uk} \affiliation{\addressNottingham}

\author{Daniel G. Figueroa} \email{daniel.figueroa@ific.uv.es} \affiliation{\addressIFIC} 
\author{Joanes Lizarraga}\email{joanes.lizarraga@ehu.eus} \affiliation{\addressEHU}\affiliation{\addressEHUQC}

\date{\today}

\begin{abstract}
We study the simultaneous decay of global string loops into scalar particles 
(massless and massive modes) and gravitational waves (GWs). Using field theory 
simulations in flat space-time of isolated loops with initial length $\sim 80-1700$ times their core width, we determine the power emitted into scalar particles, $P_{\varphi}$, and GWs, $P_{\rm GW}$, and characterize the loop decay timescale as a function of its initial length, energy and angular momentum. We quantify infrared and ultraviolet lattice dependencies of our results. For all type of loops and initial conditions 
considered, GW emission is always suppressed compared to particles as $P_{\rm GW}/P_{\varphi} \approx \mathcal{O}(10)(v/m_\text{p})^2\ll 1$, where $v$ is the vacuum expectation value associated with string formation. These conclusions are robust for the length-to-width ratios considered, with no indication they should change if the ratio is increased. The results suggest that the GW background from a global string network, such as in dark matter axion scenarios, will be suppressed compared to previous expectations.
\end{abstract}


\maketitle

\section{Introduction}
Cosmic string networks~\cite{Kibble:1976sj} 
are predicted by a variety of field theory and superstring 
early-Universe scenarios
~\cite{Kibble:1976sj,Kibble:1980mv,Vilenkin:1984ib,Hindmarsh:1994re,Copeland:2009ga,Copeland:2011dx,Vachaspati:2015cma}. They consist  of `long' strings stretching across the observable universe and string loops. The energy scale at which the network forms determines the string tension $\mu$. As the network evolves, the long-string density decreases as they intercommute forming loops. In the Nambu-Goto (NG) approximation of infinitely thin strings, loops decay mainly into gravitational waves (GWs), leading to a stochastic  
GW background (GWB)~\cite{Vilenkin:1981bx,Hogan:1984is,Vachaspati:1984gt}. 

Cosmic strings create anisotropies in the Cosmic Microwave Background (CMB)~\cite{Ade:2013xla,Lizarraga:2014xza,Charnock:2016nzm,Lizarraga:2016onn,Lopez-Eiguren:2017dmc}, and exhibit a variety of other potentially observable effects, such as non-Gaussianity in the CMB~\cite{Figueroa:2010zx,Ringeval:2010ca,Regan:2014vha}, lensing events~\cite{Vilenkin:1984ea,Bloomfield:2013jka}, and cosmic rays from the decay of strings into particle radiation~\cite{Brandenberger:1986vj,Srednicki:1986xg,Bhattacharjee:1991zm,Damour:1996pv,Wichoski:1998kh,Peloso:2002rx,Sabancilar:2009sq,Vachaspati:2009kq,Long:2014mxa,Auclair:2019jip}. The stringent constraints on the strings' tension comes however from direct searches of the cosmic string GWB. As a matter of fact, while LIGO and VIRGO have placed strong constraints on the cosmic string GWB at $\sim10-100$ Hz frequencies~\cite{Abbott:2017mem,LIGOScientific:2021nrg}, pulsar timing array (PTA) collaborations~\cite{NANOGrav:2023gor, Antoniadis:2023ott,Reardon:2023gzh, Xu:2023wog} have recently announced the first evidence for a GWB at the $\sim\text{nHz}$ frequency window. Although a GWB from a population of supermassive black hole binaries (SMBHBs) is expected at these frequencies~\cite{NANOGrav:2023hfp,Antoniadis:2023xlr}, cosmological backgrounds also represent a viable explanation~\cite{NANOGrav:2023hvm,Antoniadis:2023xlr,Figueroa:2023zhu}, in particular, a cosmic (super)string~\cite{Copeland:2003bj,Copeland:2009ga} GWB~\cite{NANOGrav:2023hvm,Antoniadis:2023xlr,Figueroa:2023zhu,Buchmuller:2023aus,Ellis:2023tsl,Wang:2023len,Kitajima:2023vre,Basilakos:2023xof,Servant:2023mwt}.  
Fitting the PTA data with a signal from NG cosmic strings leads to a tight constraint on the string tension, 
$\sqrt{\mu} \approx 1.32^{+0.20}_{-0.24} \times 10^{14}~{\rm GeV}$~\cite{Figueroa:2023zhu}, see also \cite{Kume:2024adn} for a weaker constraint when the evolution of Abelian-Higgs strings is considered.
The fit to the data is however not as good as for realistic population of SMBHBs or other cosmological signals, such as cosmic superstrings~\cite{NANOGrav:2023hvm,Antoniadis:2023xlr,Figueroa:2023zhu}. 
The potential detection of a signal from string networks in the mHz-kHz window by next generation GW detectors like LISA~\cite{Audley:2017drz} 
and others~\cite{Caprini:2018mtu,Gouttenoire:2019kij},  
with string tensions down to 
$\sqrt{\mu}\gtrsim 10^{10}$ GeV~\cite{Auclair:2019wcv}, is also an exciting prospect.

Given these developments, revisiting the GW-signal calculation from cosmic strings seems in order. {\it A crucial aspect which remains to be resolved is to understand whether GWs are in fact the key decay channel of cosmic strings.} 
Being primarily field theory objects, a natural 
decay route is particle production, which has been argued to be the primary decay route for Abelian-Higgs strings~\cite{Hindmarsh:2017qff,Hindmarsh:2021mnl}.
Recently, \cite{Matsunami:2019fss} 
set up and evolved Abelian-Higgs loop configurations, comparing the observed particle production 
with the traditional GW result from NG loops. For loops below a critical length, they found decay primarily through particle production, whilst for larger loops GW emission dominates. In~\cite{Saurabh:2020pqe} they extended their work to global strings, considering  the decay into massless and massive modes. 

In this work we further extend the approach in~\cite{Saurabh:2020pqe} with one important addition: a true comparison of the relative decay into particles and GWs requires both processes to happen simultaneously. Here we do just that, using lattice simulations of global loops created from networks as in~\cite{Hindmarsh:2019csc, Hindmarsh:2021vih} or following the methodology of~\cite{Saurabh:2020pqe}. 
Our main result is that the power emitted into GWs by isolated loops, $P_\text{GW}$, is in all cases suppressed compared to that of particle production, $P_\varphi$, as $P_{\rm GW}/P_{\varphi} \approx \mathcal{O}(10)(v/m_\text{p})^2\ll 1$, with $v$ the vacuum expectation value ($vev$)  
associated to string formation and $m_\text{p}\approx 2.44\times 10^{18}$ GeV the reduced Planck mass. This result is found to hold independently of the initial length, energy and angular momentum of the loops, and is robust for the length-to-width ratios considered, which range as $\sim 80-1700$, with no indication it should change if the ratio is increased. While we focus on global string loops here, an analogous study for Abelian-Higgs strings will appear shortly.

This paper is organized as follows. In Sec.~\ref{sec:model} we describe the model of study, how the different types of loops are generated and explain the main observables we measure. The results for the decay into particles and GWs is presented in Sec.~\ref{sec:results}. We finalize with some conclusions in Sec.~\ref{sec:conclusion}. The paper also contains one appendix, App.~\ref{app:simulation}, which summarizes the parameters of the simulations presented. 


\section{Model and loop configurations}\label{sec:model}
We consider a model with a complex scalar field $\varphi=(\phi_1+i\phi_2)/\sqrt{2}$ and action
\bea\label{eq:action}
S = -\int \text{d}^4 x \sqrt{-g}\left[\partial_{\mu}\varphi^*\partial^{\mu}\varphi + \lambda\left(\varphi\varphi^*-\frac{v^2}{2}\right)^2 
\right],
\eea
with parameters $\lambda$ and $v$ of mass-dimension 0 and 1, respectively. This model 
exhibits two phases: a symmetric phase with $\langle\varphi\rangle=0$, and a broken phase with $\langle \varphi^*\varphi\rangle=v^2/2$, where both 
massless ($\theta$) and massive ($\chi$) excitations of mass $m_\chi=\sqrt{2\lambda}v$, are present. When a transition from the unbroken to the broken phase takes place, global cosmic strings form. These are line-like topological defects with long-range interactions and a core of radius $r_\text{c} \sim 1/m_\chi$ trapped in the symmetric phase. 

In this work, we study the decay of global string loops in flat space-time using lattice simulations performed with $\mathcal{C}$osmo$\mathcal{L}$attice~\cite{Figueroa:2020rrl, Figueroa:2021yhd}. The GW emission is obtained with the {\tt GW module} of $\mathcal{C}$osmo$\mathcal{L}$attice~\cite{GWmodule:2023}, based on the algorithm introduced in~\cite{Garcia-Bellido:2007fiu}. 
We use 
lattices with periodic boundary conditions, side length  $L$,  $N$ sites per dimension and lattice spacing $\delta x = L/N$. We consider two types of loops, network and artificial loops, which are generated using different procedures, that we now describe.

\subsection{Network loops} 

\begin{figure*}[t]
  \includegraphics[width=0.30\textwidth]{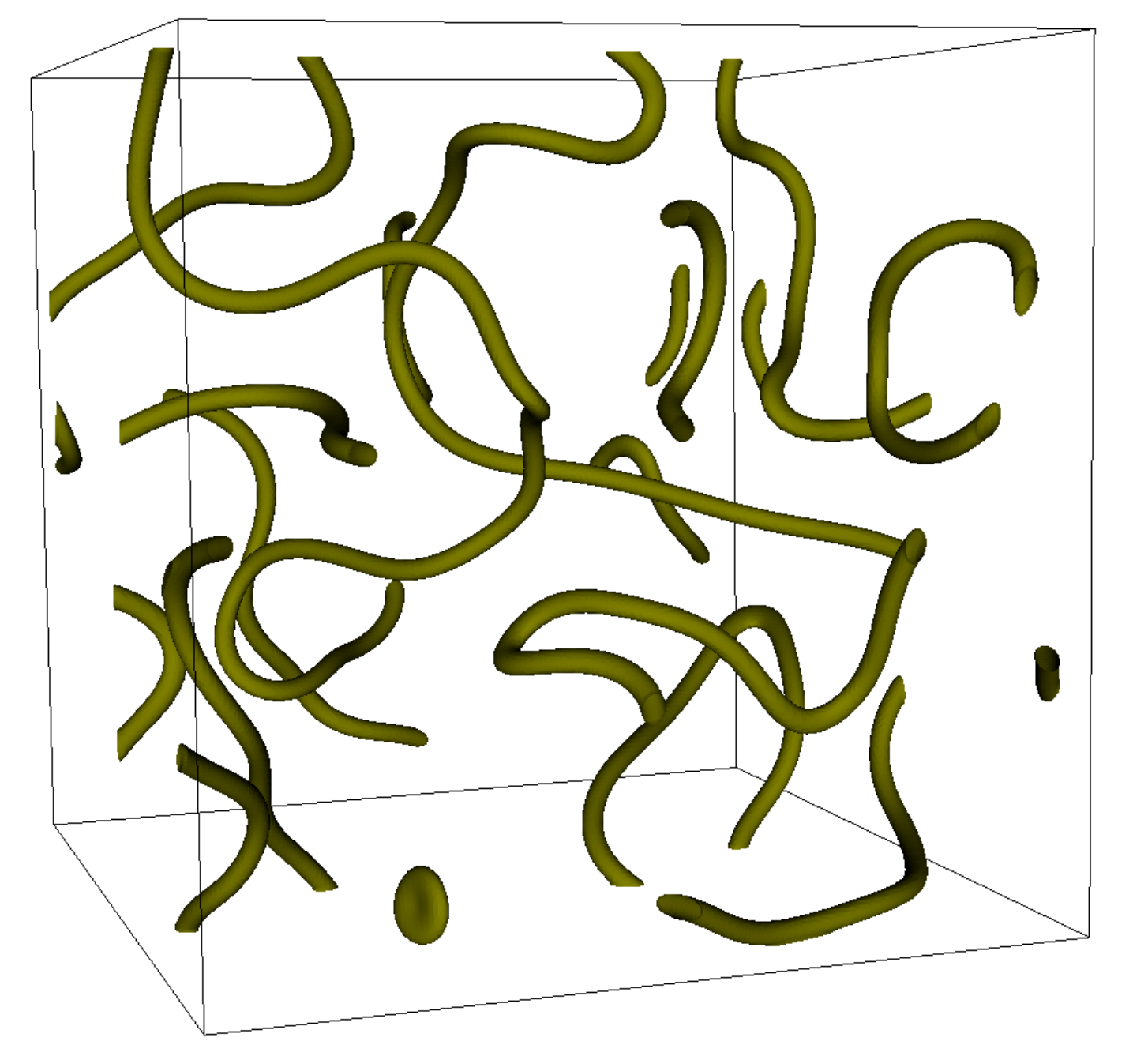}
  \includegraphics[width=0.30\textwidth]{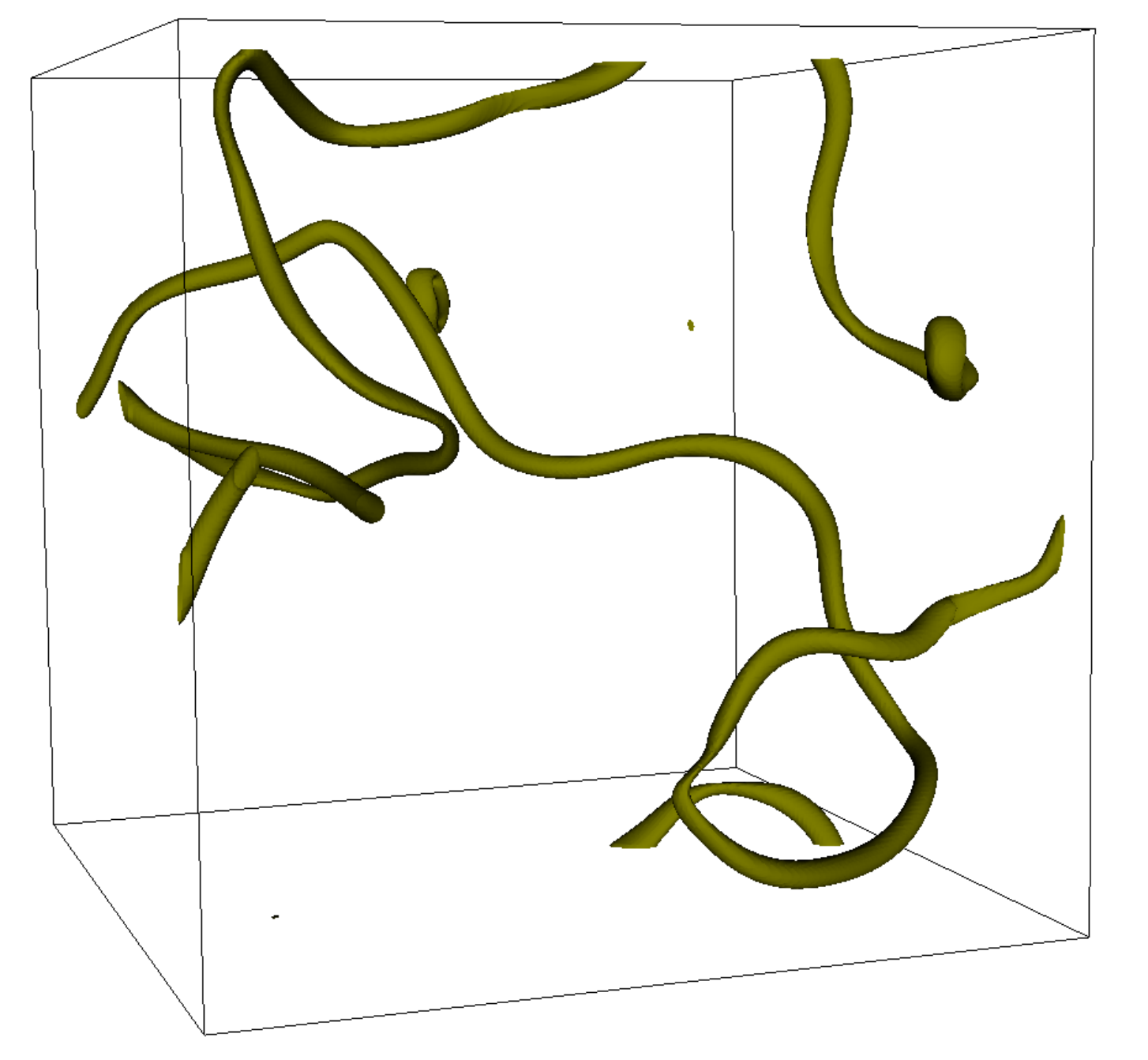}
  \includegraphics[width=0.30\textwidth]{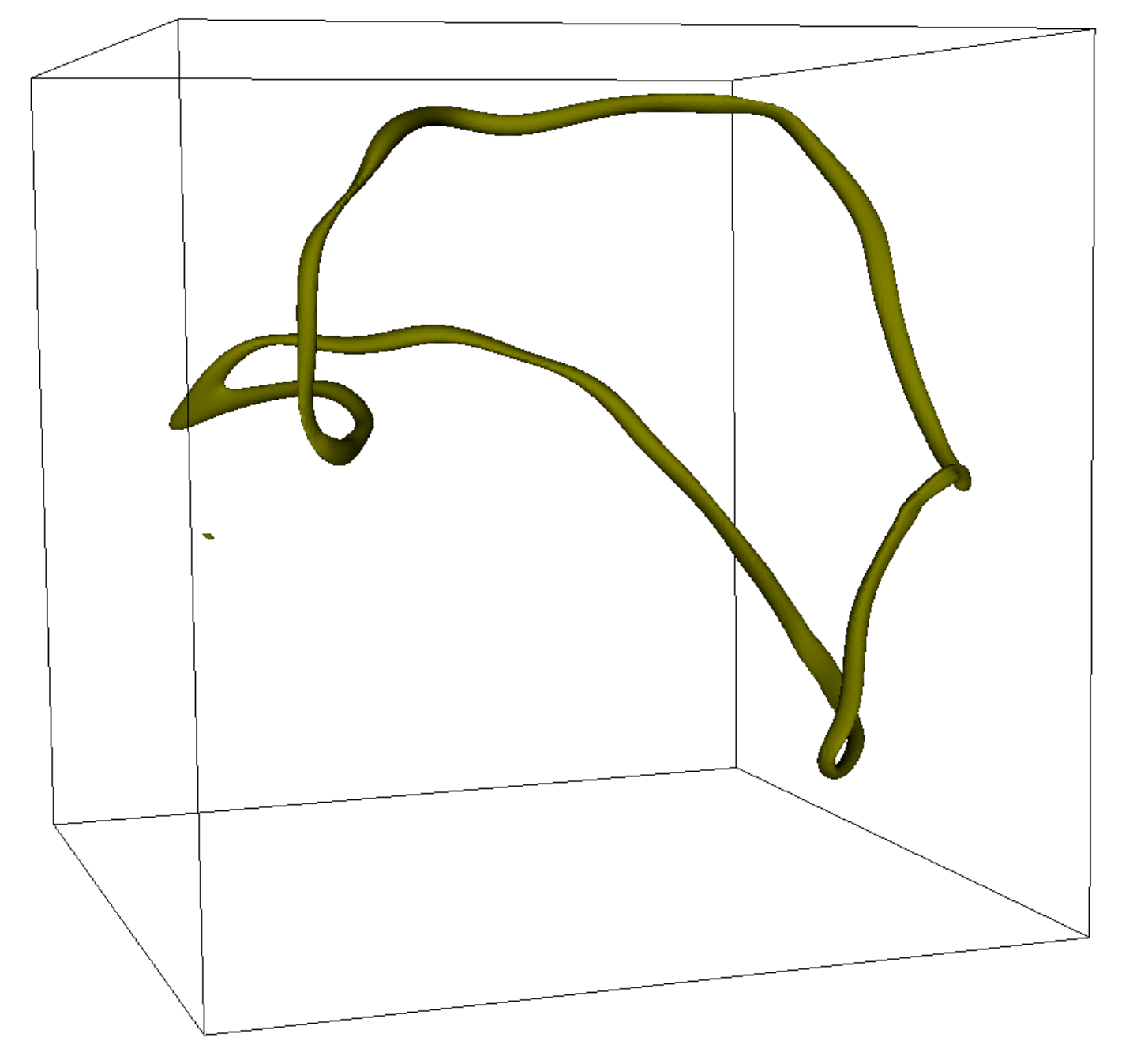}
     \caption{Three-dimensional snapshots of $|\varphi|^2=0.1v^2$ surfaces from a network simulation with $\tilde{L}=64$ and $\delta \tilde{x}=0.25$. The snapshots correspond to the end of diffusion at $\tau=\tau_0=70$ (left), the end of the extra-fattening phase at $\tau=84.5$ (center), and the detection of a single isolated loop at $\tau=96$ (right), with $\tau=\sqrt{\lambda}v t$ the program time.}
     \label{fig:networkevolution}
 \end{figure*}

Network loops originate from the decay of string networks which are close to the scaling regime~\cite{Vilenkin:1982ks,Baier:1985cn,Martins:1996jp}. These networks are generated following the procedure detailed in~\cite{Hindmarsh:2021vih}. Namely, we start simulations with a realization of a random Gaussian field, $\varphi = (\phi_1 + i \phi_2)/\sqrt{2}$, in Fourier space, with power spectrum for each field component 
\begin{equation}\label{eq:initialPS}
\mathcal{P}_{\phi_i}(k)=\frac{k^3 v^2 \ell_\text{str}^3}{\sqrt{2\pi}}\exp\left(-\frac{1}{2}k^2\ell_\text{str}^2\right)\,,\quad i = 1,~2\,,
\end{equation}
normalized such that $\langle \phi_1^2 + \phi_2^2\rangle =v^2$, where $\langle\cdots\rangle$ denotes expectation value. Here $\ell_\text{str}$ is a correlation length that controls the initial string density of the network. The resulting field configuration is initially too energetic, so we remove the excess energy by evolving the fields through a diffusion process, modeled via the equation
\begin{equation}
\dot{\phi}_i-\nabla^2\phi_i=-\lambda\left(\phi_1^2+\phi_2^2-v^2\right)\phi_i\,, 
\end{equation}
where $i=1,2$ and $\dot{\phi}_i \equiv d\phi_i/dt$.
We diffuse the field for 20 units of program time (defined below), which is enough to leave a smooth string configuration. An example of the resulting string network is represented in the left panel of Fig.~\ref{fig:networkevolution}.

Following the diffusion period, the string network is evolved in a radiation dominated (RD) background, with equation of motion
\begin{equation}\label{eq:EOM}
\ddot{\phi}_i+2\frac{\dot{a}}{a}\dot{\phi}_i-\nabla^2\phi_i=-a^2\lambda\left(\phi_1^2+\phi_2^2-v^2\right)\phi_i\,,
\end{equation}
where $i=1,2$, $t$ denotes now the conformal time  with $t_0$ the value when evolution begins, and $a\equiv a(t) =t/t_0$ is the scale factor in RD. In our simulations, we use $t_0=70/\sqrt{\lambda} v$. The background expansion is maintained for a half-light-crossing time of the lattice, $\Delta t_\text{HL}=L/2$. To avoid losing resolution of the string cores due to the expansion of the universe, the string-core resolution-preserving approach from \cite{Press:1989yh} is adopted, also known as {\it extra-fattening}, in which the coupling parameter is promoted to a time-dependent constant, $\lambda\rightarrow a^4\lambda$. This extra-fattening phase lasts for a total time $\Delta t_\text{ef}=\sqrt{t_0(t_0+\Delta t_\text{HL})}$, so that at time $t_0+\Delta t_\text{HL}$ the string-core width is equal to the one at the end of diffusion. The middle panel of Fig.~\ref{fig:networkevolution} shows an example of the network at the end of the extra-fattening phase. 

After evolving the network in an expanding background for a time $\Delta t_\text{HL}$, that includes the extra-fattening phase, networks are close to the scaling regime, with the mean string separation growing almost linearly in conformal time and the mean square velocity being constant. However, the networks have not yet decayed into an isolated loop. We then evolve the field configuration in a Minkowski background ($a = 1, \dot a = 0$) for a maximum time of $\Delta t_\text{HL}/2$, which we find to be long enough for the networks to decay into a single loop. An example of such a loop is shown in the right panel of Fig.~\ref{fig:networkevolution}. We find that $\sim 35\%$ of the simulations have decayed into single loops after this time, with the remaining forming multiple loops or infinite strings, and hence not being suitable for our study. 

After finding an isolated loop, we turn on the emission of GWs and study the evolution of the loop, still in Minkowski space-time, until 
it completely decays. It turns out that only $\sim40\%$ of the isolated loops could be used for this study. We discarded those for which the loop either self-intersected forming multiple loops of comparable size, or became infinite strings which wrapped itself around the box (thereby preventing its decay), or was initially much longer than the box size.

\subsection{Artifical loops}

Artificial loops are generated from the intersection of two pairs of boosted parallel infinite strings, following the procedure proposed in~\cite{Saurabh:2020pqe}.  We consider lattice coordinates varying from 0 to $L$, where $L$ is the lattice length side, and take the two pairs to be parallel to the $z$ and $x$ axis, respectively. Quantities associated to the pair parallel to the $z$ axis will be denoted with a ``1'' subscript, and those to the pair parallel to the $x$ axis with a ``2'' subscript.

We first describe how the  pair parallel to the $z$ axis is produced. To generate each of the strings, we consider the Nielsen-Olsen (NO) vortex solution for an infinite string, $\varphi_\text{NO}^{(w)}(x,y)=f(r)~\text{exp}(w i\theta)/\sqrt{2}$, where $(r,\theta)$ are cylindrical coordinates around the $z$-axis, $f(r)$ represents the radial profile of the vortex, and $w = +$ or $w = -$ is just a sign that specifies the {\it winding} orientation. A static string solution (of either winding) can be boosted with velocity ${\vec v}_1=v_1(\sin \alpha_1, \cos \alpha_1)$ in the $(x,y)$-plane, as $\varphi_{{\vec v}_1}^{(\pm)}(x,y;t)=\varphi_\text{NO}^{(\pm)}(x',y')$, where
\begin{eqnarray}
   x'&=-\gamma_1 v_1 s_1 t + [1+(\gamma_1-1)s_1^2] x+(\gamma_1-1)s_1c_1y\,,\\
   y'&= -\gamma_1 v_1 c_1 t + (\gamma_1-1)s_1c_1x+[1+(\gamma_1-1)c_1^2]y\,, 
\end{eqnarray}
with 
$c_1=\cos\alpha_1$, $s_1=\sin\alpha_1$ and $\gamma_1=(1-v_1^2)^{1/2}$.

\begin{figure*}[t]
\includegraphics[width=0.3\textwidth]{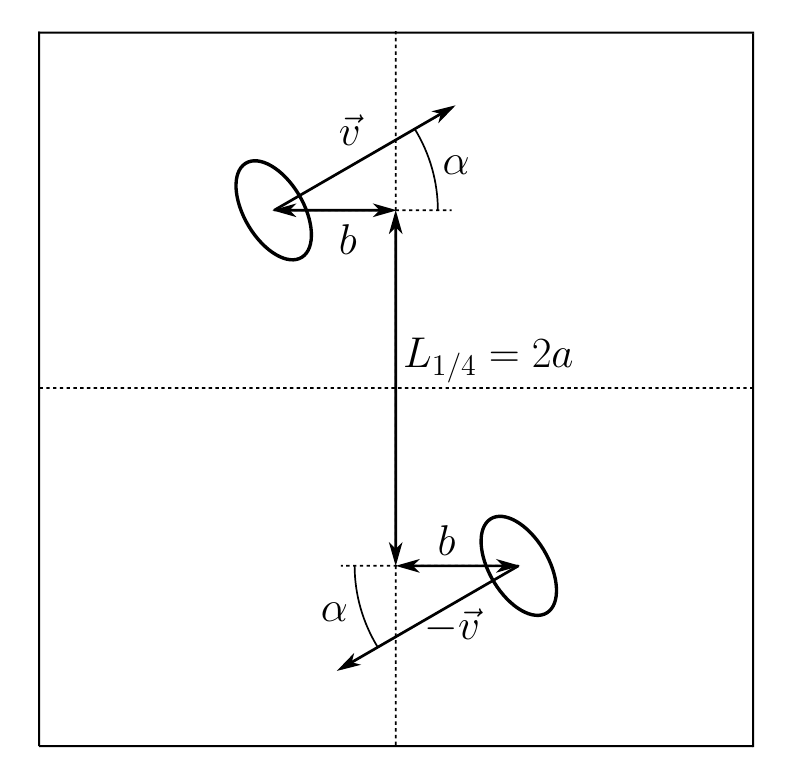}\hspace{0.1\textwidth}
\includegraphics[width=0.3\textwidth]{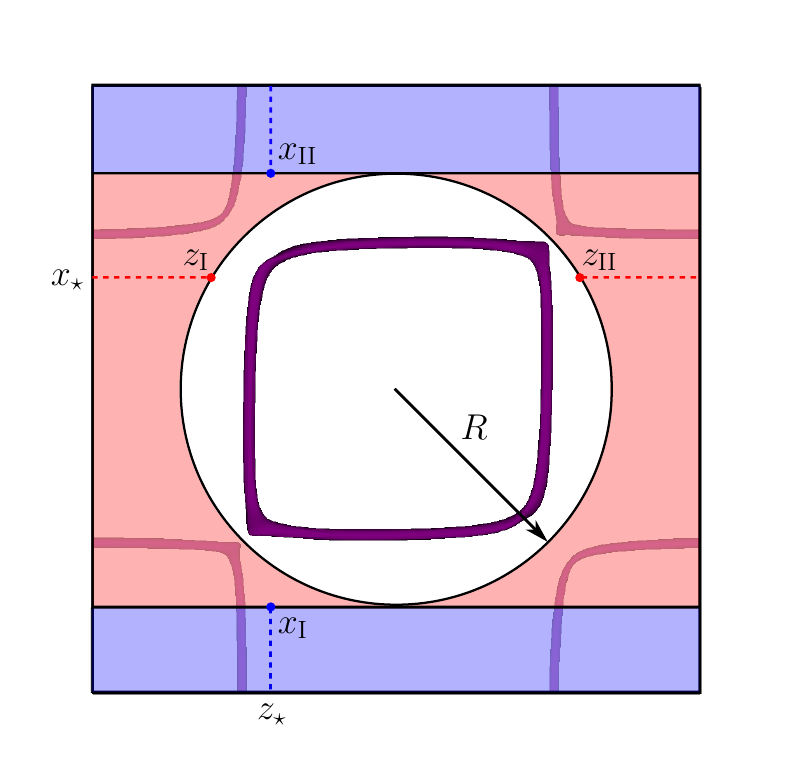}
     \caption{Schematic representation of the relevant variables used for the creation of a pair of parallel strings (left) and for isolating the inner loop (right). In the left panel, $b$ has been exaggerated compared to the actual simulations, which use $r_\text{c}\lesssim b \ll a = L_{1/4}/2$. In the right panel, red and blue regions are substituted by a smooth field configuration in two separate steps of the isolation procedure.}
     \label{fig:initialization}
 \end{figure*}

This allows us to obtain the field and its time-derivative. A pair of strings parallel to the $z$ axis is then constructed using the product ansatz~\cite{Saurabh:2020pqe} on two displaced strings with opposite windings and boosted with opposite velocities, ${\vec v}_1$ and $-{\vec v}_1$, as
\begin{multline}
    \varphi_1(x,y;t)=\frac{1}{v}\varphi_{{\vec v}_1}^{(+)}\left[x-\left(\frac{L}{2}+a_1\right),y-\left(\frac{L}{2}+b_1\right);t\right] \\ \times \varphi_{-{\vec v}_1}^{(-)}\left[x-\left(\frac{L}{2}-a_1\right),y-\left(\frac{L}{2}-b_1\right);t\right]\,,
\end{multline}
where $a_1$ and $b_1$ refer to the displacement of the strings in the $x$ and $y$ axis with respect to the center of the box, respectively, see left panel of Fig.~\ref{fig:initialization}. 
Note that we consider equal velocity magnitudes and displacements for both of the strings of each pair.

\begin{figure*}[t]
\includegraphics[width=0.30\textwidth]{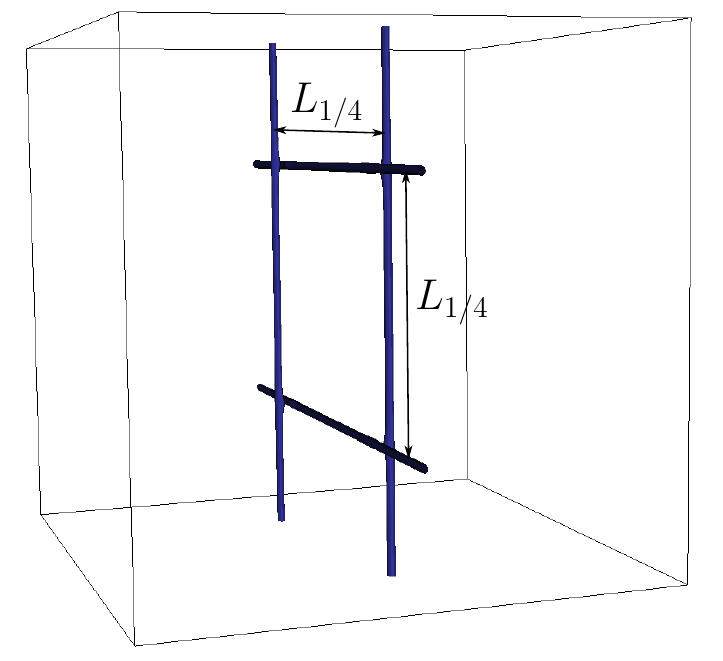}
\includegraphics[width=0.30\textwidth]{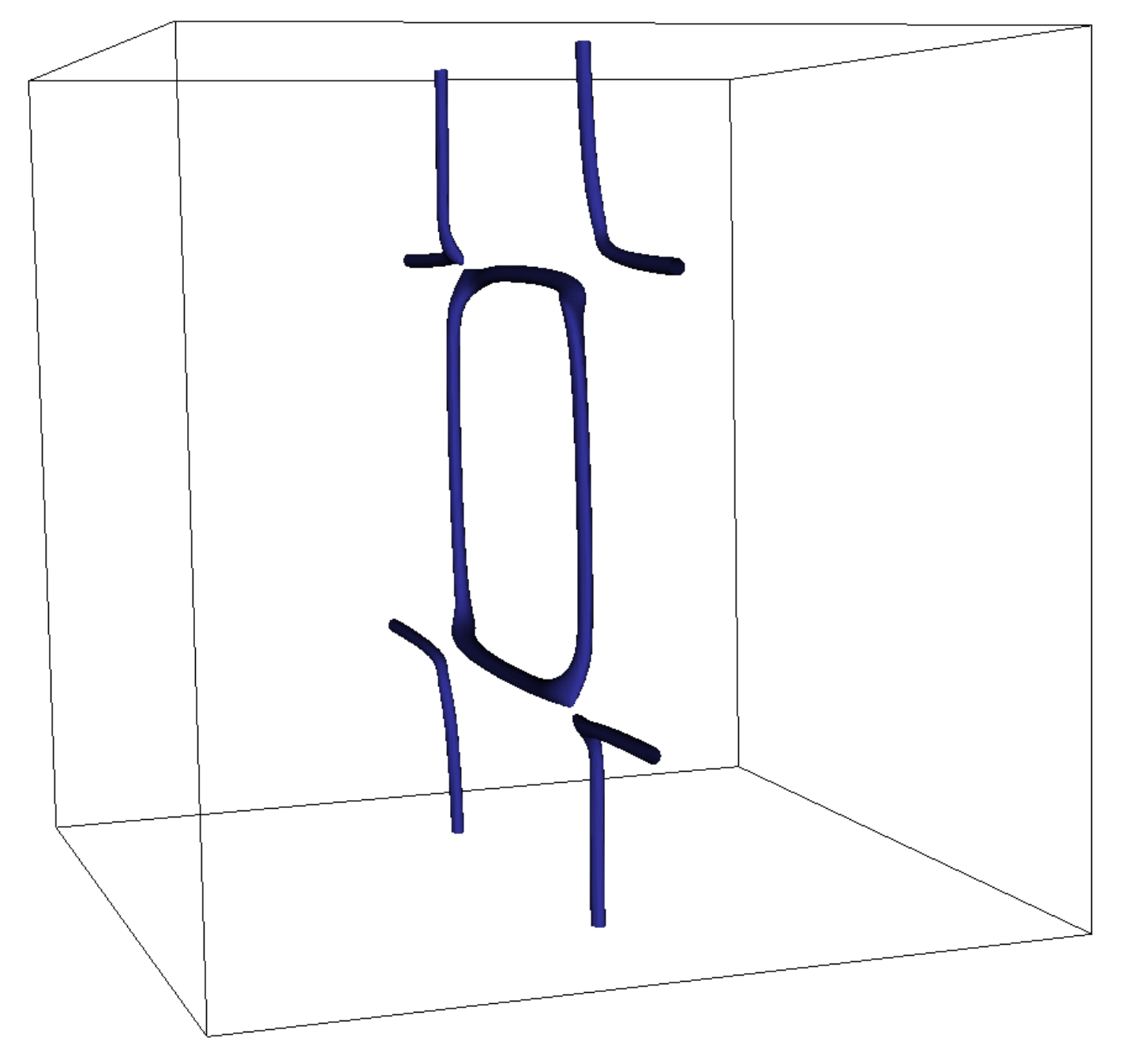}
\includegraphics[width=0.30\textwidth]{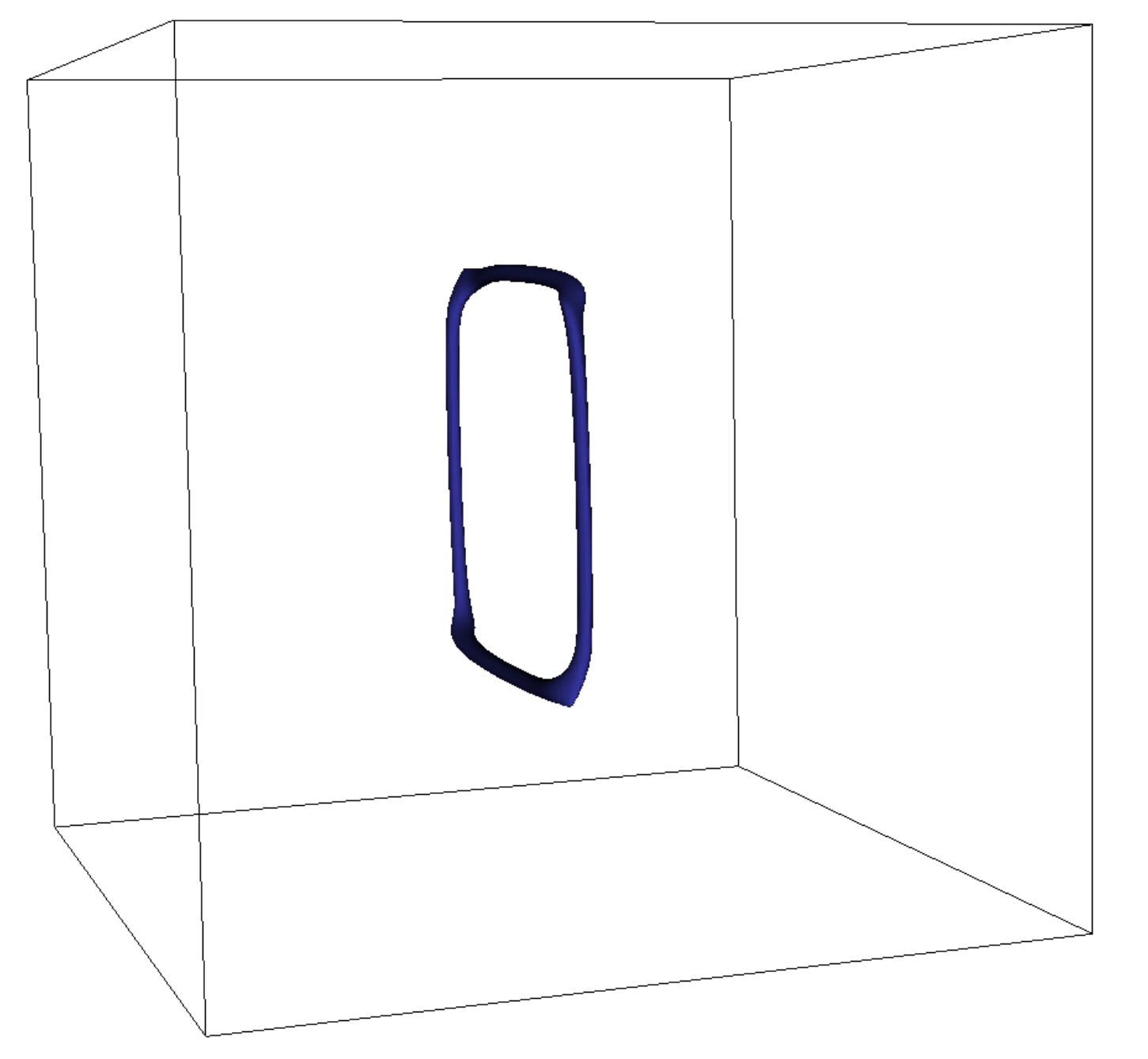}
     \caption{Three-dimensional snapshots of the $|\varphi|^2=0.06v^2$ surfaces from an artificial loop simulation with  initial conditions $v_1=0.8, v_2=0.6$, $ \sin\alpha=0.5$ and simulation parameters $\tilde{L}=64$, $L_{1/4}=32$ and $\delta \tilde{x}=0.25$. The snapshots correspond to the start of the simulation at $\tau=0$ (left), and to times immediately before (center) and after (right) the application of the isolation procedure at $\tau=2.5$, with $\tau=\sqrt{\lambda}v t$ the program time.}
     \label{fig:artificialevolution}
 \end{figure*}

The resulting field and its time derivative, evaluated at $t=0$, are modified  following~\cite{Saurabh:2020pqe} to incorporate them in a periodic lattice. Finally the 
the complete initial configuration is obtained by multiplying two pair ansatzes, each  parallel to the $z$ and $x$ axis, respectively, 
\begin{eqnarray}
\varphi(x,y,z) \equiv \varphi_1(x,y;t=0) \times \varphi_2(z,y;t=0) \,.   
\end{eqnarray} 
An example of the resulting field configuration is shown in the left panel of Fig.~\ref{fig:artificialevolution}.

In this work, we consider each of the string pairs to have, in general, different velocity magnitudes, $v_1\neq v_2$ with $v_2\geq v_1$, but the same angle with respect to the normal to the plane where the strings intersect, i.e.~$\alpha_1=\alpha_2=\alpha$. We also consider all strings to initially lie almost in that same plane. We let $a_1=a_2$ be a significant fraction of the lattice size, typically $L/4$ or $L/6$, while $b_1=0$ and $b_2=2/\sqrt{\lambda} v$. This choice ensures 
strings are enough separated so that 
the product ansatz remains valid, while the two pairs are close enough so that the intersection happens very early in the simulation. The (approximate) separation of two parallel strings within each pair is denoted $L_{1/4}=2a$.

The initial configuration is evolved using Eq.~(\ref{eq:EOM}) in a flat Minkowski background ($a = 1, \dot a = 0$). The four straight strings rapidly intersect forming two square-shaped configurations, an {\it inner} loop of initial length $L_0\approx 4L_{1/4}$, and an {\it outer} loop. Shortly after, the two loops start to shrink (due to particle emission) and separate from each other. As we expect the inner loop to be less affected by the initial condition than the outer loop, we have developed a procedure to isolate the inner loop once both loops are sufficiently away from each other. The procedure is based on the fact that both loops are almost planar and the center of the inner loop is close to the center of the box, see right panel of Fig.~\ref{fig:initialization}.

The isolation procedure consists on the following. After formation, we let the two loops evolve until their separation is a small fraction (we choose a factor $0.15$) of the radius of the inner loop. At this point, we consider a cylinder of radius $R$ and axis parallel to the $y$ axis, such that it encompasses the inner loop and its surface lies halfway between the two loops. The field outside of the cylinder and its time derivative are substituted with a smooth configuration, removing the outer loop. The substitution is performed in two steps, repeated for each fixed value of the coordinate $y=y_\star$. First, we consider the points $(x,y_\star,z)$ outside the cylinder with $|L/2-x| < R$ (red area in the right panel of Fig.~\ref{fig:initialization}). Each possible value of $x=x_\star$ defines a line that intersects the cylinder at $z_\RNum{1}$ and $z_\RNum{2}>z_\RNum{1}$. The field and its time derivative outside the cylinder in this line are substituted with a smooth configuration. The phases are linearly interpolated using the phase values at $z_{\RNum{1},\RNum{2}}$, while for the modulus an ansatz based on the long-distance behavior of the NO solution is used,
\begin{widetext}
\begin{equation}
\begin{array}{rll}
\displaystyle |\varphi(x_\star,y_\star,z)|&=\left\{\begin{array}{l}
\displaystyle\frac{\xi_\RNum{1}}{(\Delta-z+z_\RNum{1})^2}+\frac{\xi_\RNum{2}}{(L + z + \Delta - z_\RNum{2})^2}+\frac{v}{\sqrt{2}}\,,\\[2ex]
\displaystyle\frac{\xi_\RNum{1}}{(L-z+\Delta+z_\RNum{1})^2}+\frac{\xi_\RNum{2}}{(\Delta+z-z_\RNum{2})^2}+\frac{v}{\sqrt{2}}\,, 
\end{array}
\right.\quad\quad &
\begin{array}{l}
 \displaystyle 0\leq z < z_\RNum{1}\,,\phantom{\frac{A_\RNum{1}}{\Delta^2}}\\
 \displaystyle z_\RNum{2}<z<L\,,\phantom{\frac{A_\RNum{1}}{\Delta^2}}
\end{array}\\[7ex]
\displaystyle |\dot{\varphi}(x_\star,y_\star,z)|&=\left\{\begin{array}{l}
\displaystyle\frac{\pi_\RNum{1}}{(\Delta-z+z_\RNum{1})^3}+\frac{\pi_\RNum{2}}{(L + z + \Delta - z_\RNum{2})^3}\,, \\[2ex]
\displaystyle\frac{\pi_\RNum{1}}{(L-z+\Delta+z_\RNum{1})^3}+\frac{\pi_\RNum{2}}{(\Delta+z-z_\RNum{2})^3}\,, 
\end{array}
\right.\quad\quad&
\begin{array}{l}
 \displaystyle 0\leq z < z_\RNum{1}\,,\phantom{\frac{A_\RNum{1}}{\Delta^2}}\\
 \displaystyle z_\RNum{2}<z<L\,,\phantom{\frac{A_\RNum{1}}{\Delta^2}}
\end{array}
\end{array}
\end{equation}
\end{widetext}
where $\xi_{\RNum{1},\RNum{2}}$ and $\pi_{\RNum{1},\RNum{2}}$ are normalization constants that ensure the configuration is periodic and matches in the surface of the cylinder, and $\Delta$ is a length parameter we set to $\Delta=2/\sqrt{\lambda} v$. In a second step, we consider points with $|L/2-x| \geq R$, and proceed analogously working with fixed $z=z_\star$ (blue regions in the right panel of Fig.~\ref{fig:initialization}). Both steps are finally repeated for all values of $y_\star$. The central and right panels of Fig.~\ref{fig:artificialevolution} show an example of a field configuration just before and after isolating the inner loop. Once the isolation procedure is complete, we turn on GWs and study the loop until it decays.

We have checked the effect of varying $\Delta$, as well as the isolation time and the size of the cylinder, finding that the results are insensitive to these changes as long as $a)$ both strings are sufficiently separated from each other, and $b)$ the cylinder surface is not too close to any of the two loops.

\begin{widetext}
\begin{table*}[t]
\renewcommand{\arraystretch}{1.3}
\centering
\begin{tabular}{|>{\centering\arraybackslash}p{4.5cm}|>{\centering\arraybackslash}p{1.5cm}|>{\centering\arraybackslash}p{1.5cm}|>{\centering\arraybackslash}p{1.5cm}|>{\centering\arraybackslash}p{1.5cm}|}
\hline
 Type of loop& $A\times 10^3$ & $B$ & $C\times 10^3$ & $D$   \\ \hline 
 Network  & $185\pm22$ & $5\pm14$ & $89\pm13$ & $-1\pm17$ \\\hline
 Artificial, $v_1=0.9, v_2=0.9$  & $571\pm10$ & $-26\pm3$ & $270\pm9$ & $-22\pm6$  \\ \hline
 Artificial, $v_1=0.9, v_2=0.6$ & $430\pm30$ & $-13\pm11$ & $223\pm14$ & $-14\pm9$   \\ \hline
 Artificial, $v_1=0.9, v_2=0.3$ & $534\pm14$ & $-23\pm5$ & $269\pm9$ & $-20\pm6$  \\ \hline
 Artificial, $v_1=0.9, v_2=0.0$ & $706\pm19$ & $-45\pm6$ &  $348\pm9$ & $-40\pm6$ \\ \hline
 Artificial, $v_1=0.6, v_2=0.6$ & $227\pm11$ & $-3\pm4$ & $125\pm5$ & $-1\pm3$ \\ \hline
 Artificial, $v_1=0.6, v_2=0.3$ & $260\pm15$ & $-9\pm5$ & $140\pm8$ & $-6\pm5$ \\ \hline
 Artificial, $v_1=0.3, v_2=0.3$ & $280\pm30$ & $-1\pm9$ & $154\pm15$ & $2\pm9$ \\ \hline
\end{tabular}
\caption{Results of linear fits $\Delta\tau_{\rm dec} = A\tilde L_{0} + B$ and $\Delta\tau_{\rm dec} = C\tilde E_{\text{str,0}} + D$, for network and artificial loops. All artificial loops are simulated with $\sin\alpha=0.4$.
} 
\label{tab:decayfits}
\end{table*}
\end{widetext}

\subsection{Measures and observables}

In order to describe the loop dynamics and energy transfer(s) we monitor several quantities. For the loop dynamics and decay characterisation, we follow the loop's length, energy~\cite{Hindmarsh:2021vih} and angular momentum~\cite{Saurabh:2020pqe}. The loop's length in the lattice frame is determined from counting the number of pierced plaquettes (taking into account the {\it Manhattan effect}~\cite{Vachaspati:1984dz,Rajantie:1998vv,Fleury:2015aca}), while its energy and angular momentum are defined, respectively, by
\bea\label{eq:energy}
E_\text{str}=\int \text{d}^3 x \,W(\varphi) \left[ \dot{\varphi} \dot{ \varphi}^* +\vec\nabla\varphi\cdot \vec\nabla\varphi^* + V(\varphi)\right]\,,\\[1ex]
\label{eq:angularmomentum}
\vec J=-\frac{1}{2}\int \text{d}^3 x \,W(\varphi) \left[ \vec x \times \big(\dot\varphi \vec\nabla\varphi^*+\dot\varphi^*\vec\nabla\varphi\big)\right]\,,
\eea
where $W(\varphi) = (V(\varphi)/W_0)\times\Theta(v^2/2-|\varphi|^2)$  is a {\it weight function}, with $W_0 = \lambda v^4 / 4$ and $\Theta(x)$ the step function.

We additionally measure the energy and spectra of the massless ($\theta$) and massive ($\chi$) modes~\cite{Saurabh:2020pqe, Drew:2019mzc, Drew:2022iqz}. For a real field $\phi$, its discrete power spectrum is 
\be
\Delta_\phi(k)=\frac{k^3}{2\pi^2}\left(\frac{\delta x}{N}\right)^3\langle|\phi(k)|^2\rangle_{\hat{\Omega}_k}\,,
\label{eq:ChiThetaPS}
\ee
with $\phi(k)$ the Fourier transform of $\phi$ and $\langle\cdots\rangle_{\hat{\Omega}_k}$ an angular averaging in $k$-space. For comparison with~\cite{Saurabh:2020pqe} we also measure the energy spectrum per linear intervals of the massless radiation, defined as,
\be
E_\theta(k)=\frac{v^2}{2}\left[ |\dot{\theta}({\vec k})|^2 + k^2|\theta({\vec k})|^2\right]\,.
\label{eq:EnergyPS}
\ee

\section{Results}\label{sec:results}

In this section, we present our results for the decay of both types of loops into particles and GWs. From now on, we express physical quantities in terms of dimensionless 
\textit{program variables}: 
 $\tilde{\varphi}=\varphi/v$, $\tilde{\vec x} =\sqrt{\lambda}v\,{\vec x}$, $\tau = \sqrt{\lambda}v\,t$,  $\kappa=k/\sqrt{\lambda}v$, ${\tilde E}_{\text{str}}=(\sqrt{\lambda}/v)\,E_{\text{str}}$ and $\tilde J = \lambda |\vec J|$.

\subsection{Loop decay}

We characterize the decay of loops into scalar particles and their decay time, $\Delta\tau_{\rm dec}$, as a function of their initial length, ${\tilde L}_0$, and other physical observables. 
We report on the lifetime of 23 network loops with $80 \lesssim \tilde{L}_0/\tilde r_{\rm c} \lesssim 1700$, where $\tilde r_{\rm c} = 1/\sqrt{2}$, and of 49 artificial loops divided into 7 sets according to their boost velocities, with $100 \lesssim \tilde{L}_0/\tilde r_{\rm c} \lesssim 800$. The simulation parameters are summarized in Tables~\ref{tab:networkIC} and~\ref{tab:artificialIC} in App.~\ref{app:simulation}.

\begin{figure*}[t!]
\includegraphics[width=1\textwidth,height=6.3cm]{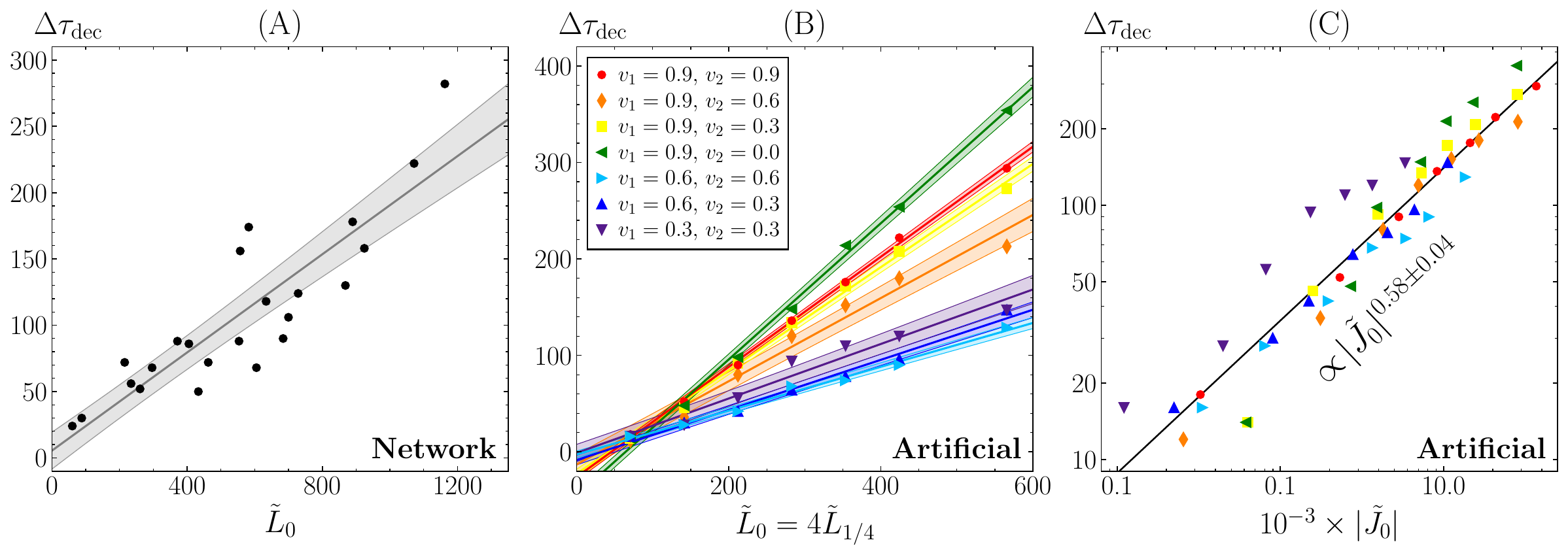}
     \caption{Lifetime as a functions of initial length for network (A) and artificial loops with different boost velocities (B), and as a function of initial angular momentum for the artificial loops (C), in a  logarithmic scale.  Lines and shaded regions in (A) and (B) correspond to linear-fit results, and to a power-law fit in (C). 
     }
     \label{fig:stringdynamics}
\end{figure*}
\begin{figure*}[t!]
\includegraphics[width=0.39\textwidth]{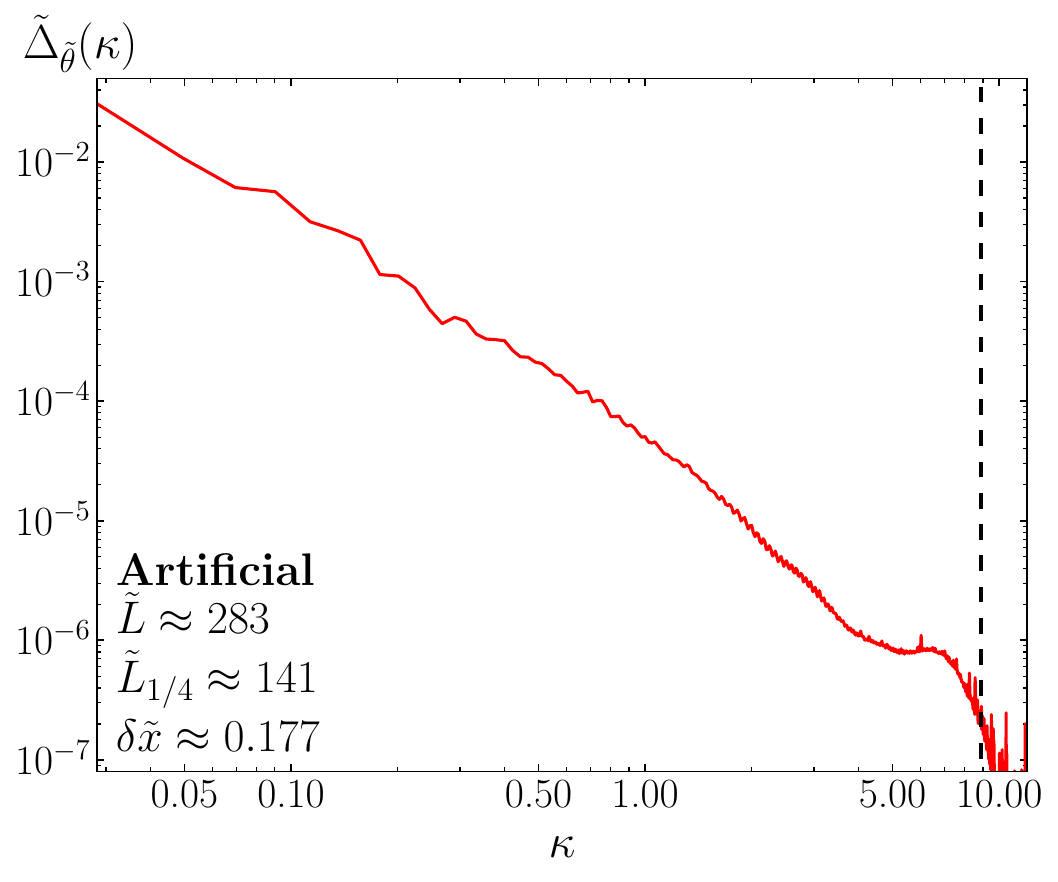}\hspace{0.1\textwidth}
\includegraphics[width=0.39\textwidth]{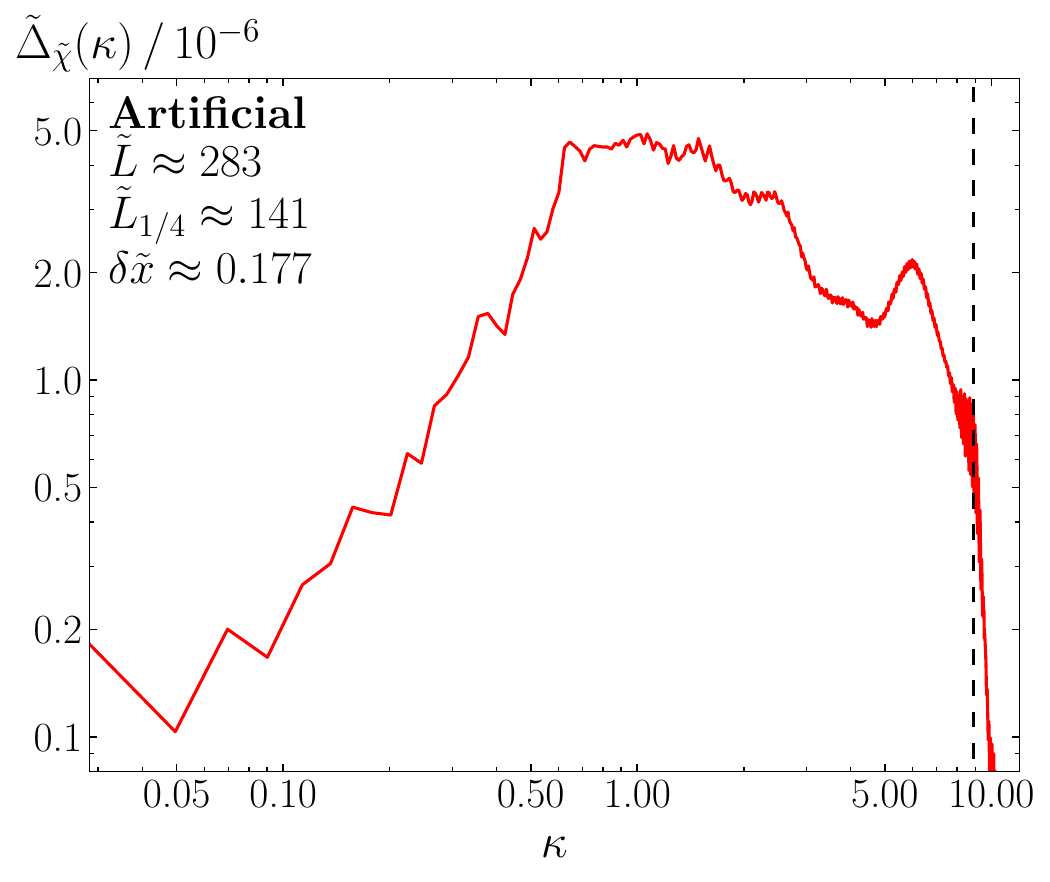}
     \caption{Power spectra of massless (left) and massive (right) particles just after the decay of an artificial loop with $v_1=v_2=0.6$ and $\sin\alpha=0.4$. Dashed vertical lines indicate the scale of the string width, $\kappa_\text{c} = 2\pi/\tilde r_\text{c} \approx 9$.}
     \label{fig:chithetaPS}
 \end{figure*}

In Fig.~\ref{fig:stringdynamics}-(A) we show $\Delta\tau_{\rm dec}$ for network loops versus $\tilde{L}_{0}$, the latter measured from the pierced plaquettes. Although the points show some scatter, $\Delta\tau_{\rm dec}$ scales roughly linearly with $\tilde{L}_{0}$,  
indicating a scale invariant mechanism driving the decay of the loop. 
A linear
fit, $\Delta\tau_{\rm dec} = A \tilde{L}_{0} + B$, is presented (solid line and band).
We find a similar fit against the initial string energy,  
$\Delta \tau_{\rm dec} = C\tilde E_{\rm str,0} + D$, 
allowing us to estimate the particle-emission power, $P_{\varphi} \equiv v^2\tilde P_{\varphi}$, as  $\tilde P_{\varphi} \equiv {\text{d}\tilde E_{\rm str,0}/\text{d}\tau} = 1/C =11.2\pm1.6$. 

In Figs.~\ref{fig:stringdynamics}-(B) and~\ref{fig:stringdynamics}-(C) 
we present $\Delta \tau_{\rm dec}$ for artificial loops versus $\tilde{L}_0$ ($\simeq 4\tilde{L}_{1/4}$), and the initial angular momentum, $\tilde J_ 0$, respectively, with each color representing a different set of velocities. Fig.~\ref{fig:stringdynamics}-(B) shows that artificial loops live longer than network loops of the same length, for the range of lengths that can be compared. Once again, linear relations 
are observed, although the  
data shows a clear dependence on the velocities, indicating that no linear scaling can fit all cases simultaneously. The particle emission power ranges between 
$\tilde P_{\varphi} \equiv {d{\tilde E}_{\rm str}/d\tau}=1/C\approx3-8$, depending on the velocity pairs. Results of all fits discussed so far, both for network and artificial loops, can be found summarized in Table~\ref{tab:decayfits}. We note that the case $v_1=v_2=0.6$ corresponds to the same setup used in~\cite{Saurabh:2020pqe}, who find the linear fit coefficient $A$ 
to be $\sim40\%$ larger than ours.

Fig.~\ref{fig:stringdynamics}-(C) shows a power-law fit to all artificial loops that roughly scales as $\Delta\tau_{\rm dec} \propto \tilde{J}_{\rm 0}^{3/5}$,  
highlighting angular momentum as a major ingredient affecting loop decay. Retrospectively, this might also explain the scatter of points in Fig.~\ref{fig:stringdynamics}-(A), since we cannot control $\tilde{J}_0$ for network loops.  


Several consistency checks have been performed to ensure robustness of our results. We observe negligible variations of $\Delta\tau_{\rm dec}$ when changing $\alpha$ or reducing the lattice spacing, and changes of less than $10\%$ in $\Delta\tau_{\rm dec}$ of artificial loops  
when increasing the ratio $\tilde L/\tilde L_{1/4}$.\\ 

\begin{center}
    {\it Particle emission}
\end{center}

To complement our results, 
we comment also on the distribution of particles produced in the decays, corresponding to both 
massless ($\theta$) and massive ($\chi$) field excitations. These are identified, respectively, with the angular and radial perturbations from the true vacuum of the theory, $\varphi=\frac{1}{\sqrt{2}}(v+\chi)\exp(i\theta)$. 

In Fig.~\ref{fig:chithetaPS} we plot as an example the power spectra of $\theta$ (left) and $\chi$ (right) [{\it c.f.~}Eq.~\ref{eq:ChiThetaPS}] at the end of the decay of an artificial loop, generated with $v_1=v_2=0.6$, $\sin\alpha=0.4$ and $\tilde{L}_{1/4} = \tilde{L}/2 \approx 141$. We note how the spectrum of the massless field is power-law suppressed at high modes, while the spectrum of the massive field peaks around $\kappa\sim 1$, albeit with an amplitude much smaller than the massless field spectrum at the same scale. The spectrum of $\theta$ reflects that {\it soft} modes ($\kappa \ll 1$) of (massless) radiation can be easily emitted, while the spectrum of $\chi$ reflects that massive mode emission is possible but it is very suppressed compared to the massless emission. 


\begin{figure}[t]
    \includegraphics[width=0.45\textwidth]{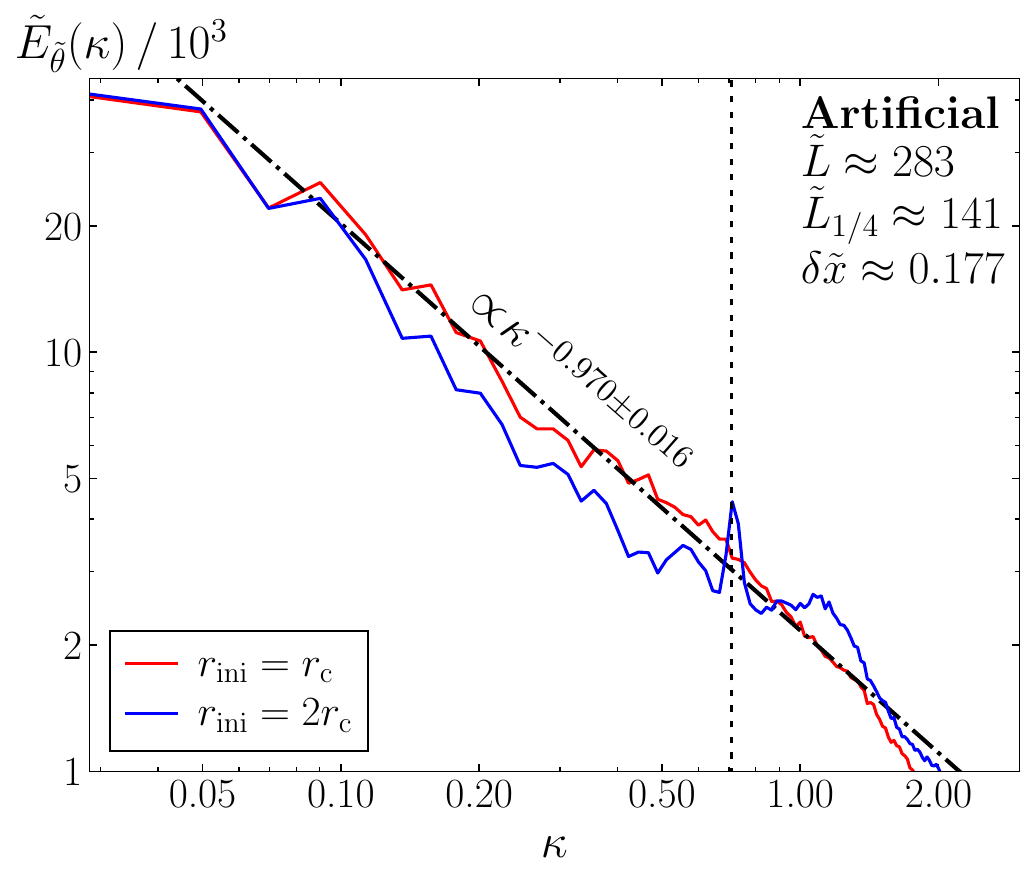} 
	\caption{Energy power spectrum per linear interval of massless radiation at the end of the decay of an artificial loop initialized with $v_1=v_2=0.6$ and $\sin\alpha=0.4$, both in the case of initial core radius, $r_\text{ini}$ equal to the physical one (red) and twice as big (blue). We observe a consistent scaling with that in \cite{Saurabh:2020pqe}, but only find the sharp peak for the physically unrealistic initial conditions.}\label{fig:PStheta}
\end{figure}

We note that~\cite{Saurabh:2020pqe} showed a peak emerging in the spectrum of $\theta$ for artificial loops, at a scale half the mass of the massive mode, $k = m_\chi/2$ ($\kappa \simeq \sqrt{2}/2 \approx 0.7$ in our units). In Fig.~\ref{fig:PStheta}, we present in red our results for the same energy spectrum (after the loop decayed), as defined in Eq.~\ref{eq:EnergyPS}. Performing a power-law fit we observe that the spectrum scales roughly as $\propto k^{-1}$, as observed in~\cite{Saurabh:2020pqe}. However, we detect no presence of a peak. Interestingly, if we set the initial radius of the string to be larger than the physical one by modifying the NO vortex solution as $f(r)\rightarrow f(2r)$, we observe a peak appearing at the same scale scale as in~\cite{Saurabh:2020pqe}. 
In our case it looks like this peak is related to an excess energy in the radial mode of the string due to an excessively large core width set in the initial strings' configuration. Furthermore, we only observe `bulges' in the three-dimensional representation of the field, as those discussed in~\cite{Saurabh:2020pqe}, when considering this extra-large initial radius.\vspace{-.04cm}

\begin{figure*}[t]
\includegraphics[width=1\textwidth,height=6.2cm]{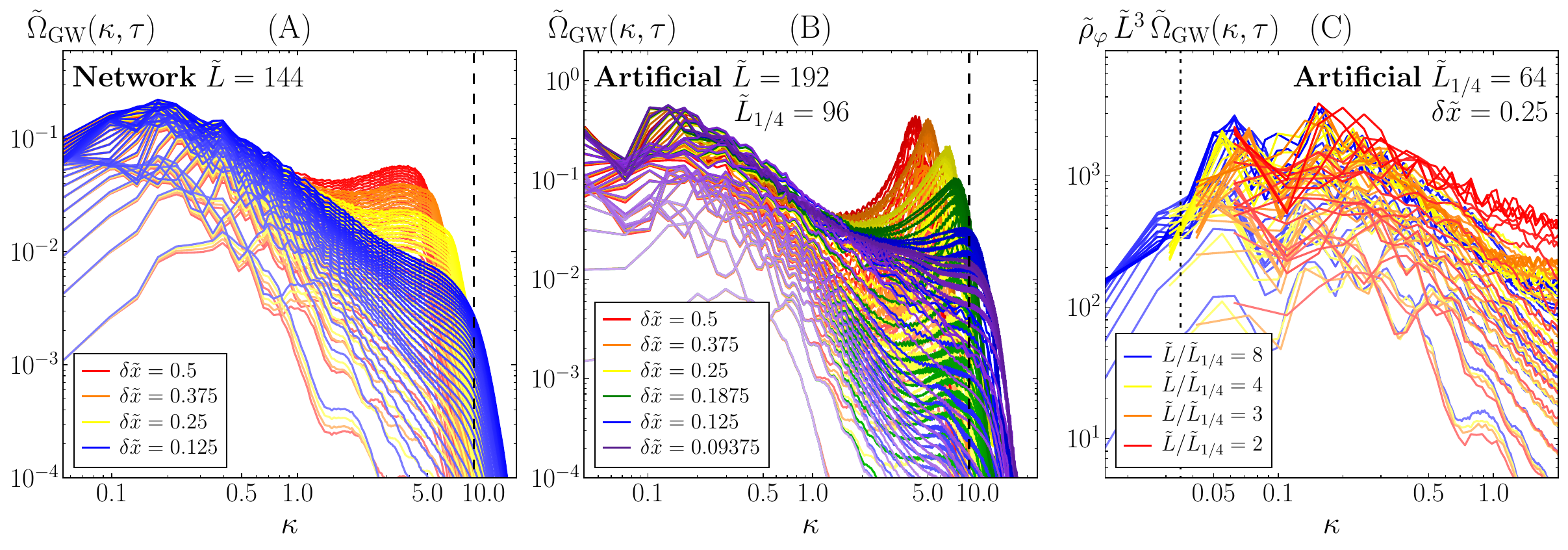}
     \caption{GW energy density spectra evolution for network (A) and artificial loops (B) with varying UV resolution, $\delta\tilde x$, and fixed lattice size, $\tilde L$, and for artificial loops with varying the $\tilde L$ and fixed $\delta\tilde x$ (C). Dashed and dotted vertical lines indicate the scale of the string width, $\kappa_\text{c} = 2\pi/\tilde r_\text{c}$, and of the initial length of the string, $\kappa_0 = 2\pi/\tilde L_0$, respectively. Spectra go from early to late times from bottom to top, with lines separated by two units of program time.}
     \label{fig:UVIRspectra}
 \end{figure*}

\subsection{GW emission} 

The GW emission from a string loop is obtained by solving 
\begin{equation}
\ddot h_{ij} - \vec\nabla^2h_{ij} = 4 m_\text{p}^{-2}[{\rm Re}(\partial_i\varphi\partial_j\varphi^*)]^{\rm TT}\,,
\end{equation}
where $h_{ij}$ are  tensor metric perturbations that represent GWs and obey the tranversality and tracelessness conditions, $\partial_i h_{ij} = h_{ii} = 0$. Here $[ \cdots ]^{\rm TT}$ implies transverse-traceless projection and $m_\text{p} \approx 2.44\times 10^{18}$ GeV is the reduced Planck mass. We work in the linear gravity regime and neglect backreaction of the GWs onto the loop dynamics, something we justify self-consistently later. The GW energy density spectrum, 
normalized to the total scalar field energy density, $\rho_{\varphi}$, is~\cite{Caprini:2018mtu}

\begin{multline}\label{eqn:GWpowerspectrum}
\Omega_{\rm GW}(k,t) \equiv  
{1\over \rho_\varphi} \frac{\text{d}\rho_{\rm GW}}{\text{d}\log k} =
\frac{k^3m_\text{p}^2}{8\pi^2L^3\rho_\varphi}\big\langle\dot{{h}}_{ij}(k,t)\dot{{h}}^*_{ij}(k,t)\big\rangle_{\hat \Omega_k},\nonumber
\hspace*{-0.5cm}\\
\end{multline}
where $\langle \cdots \rangle_{\hat \Omega_k}$ represents angular averaging in $k$-space. The total GW energy emitted by a loop is obtained via 
\begin{equation}
    E_\text{GW}(t) = \rho_{\varphi}L^3\int \Omega_{\rm GW}(k,t)\,\text{d}\log k\,.
\end{equation}
In terms of program variables, we write these expressions as
$\tilde E_{\rm GW} = (\sqrt{\lambda}/v)\,E_{\rm GW}\,/\,(v/m_\text{p})^2$, and $\tilde\Omega_{\rm GW}(\kappa,\tau) = \Omega_{\rm GW}(k,t)\,/\,(v/m_\text{p})^2$. 

We first analyze  lattice discretization effects on the GW spectrum emitted by a loop.
For network loops, we run a high resolution simulation with $\delta \tilde{x}=0.125$, and create new lattices with lower resolution, $\delta \tilde x^{(p)} = p\delta \tilde x$ ($p=2,3,4$), by eliminating $p-1$ sites (per dimension) of every $p$ consecutive points of the original lattice. For artificial loops, we run simulations with different $\delta\tilde{x}$ and fixed $\tilde{L}=192$, $\tilde{L}_{1/4}=96$, $v_1=0.6$, $v_2=0.7$ and $\sin\alpha=0.5$. Figs.~\ref{fig:UVIRspectra}-(A) and~\ref{fig:UVIRspectra}-(B) show the evolution of the GW spectrum for these network and artificial loops, respectively. In all cases, GW emission peaks at IR scales $\kappa \sim (2-6)\kappa_0$, with $\kappa_{0} = 2\pi/\tilde L_0$ the scale of the initial string length, and there is good agreement between spectra up to a scale $\kappa \sim 0.1\kappa_\text{c}$ , with $\kappa_\text{c} \equiv 2\pi/\tilde r_\text{c}$ the scale of the core radius (black dashed line). A second peak emerges at higher modes, but its height is suppressed as the UV resolution is improved, indicating it is 
a lattice artifact arising when the string core is not well resolved. We thus compute the total energy emitted in GWs by integrating the spectrum up to $\kappa_{\rm cut}$, guaranteeing independence from $\delta\tilde x$. 
Note that the GW power is completely suppressed for scales smaller than the core radius, $\kappa > \kappa_\text{c}$.

\begin{figure*}[t]
\includegraphics[width=1\textwidth,height=6.2cm]{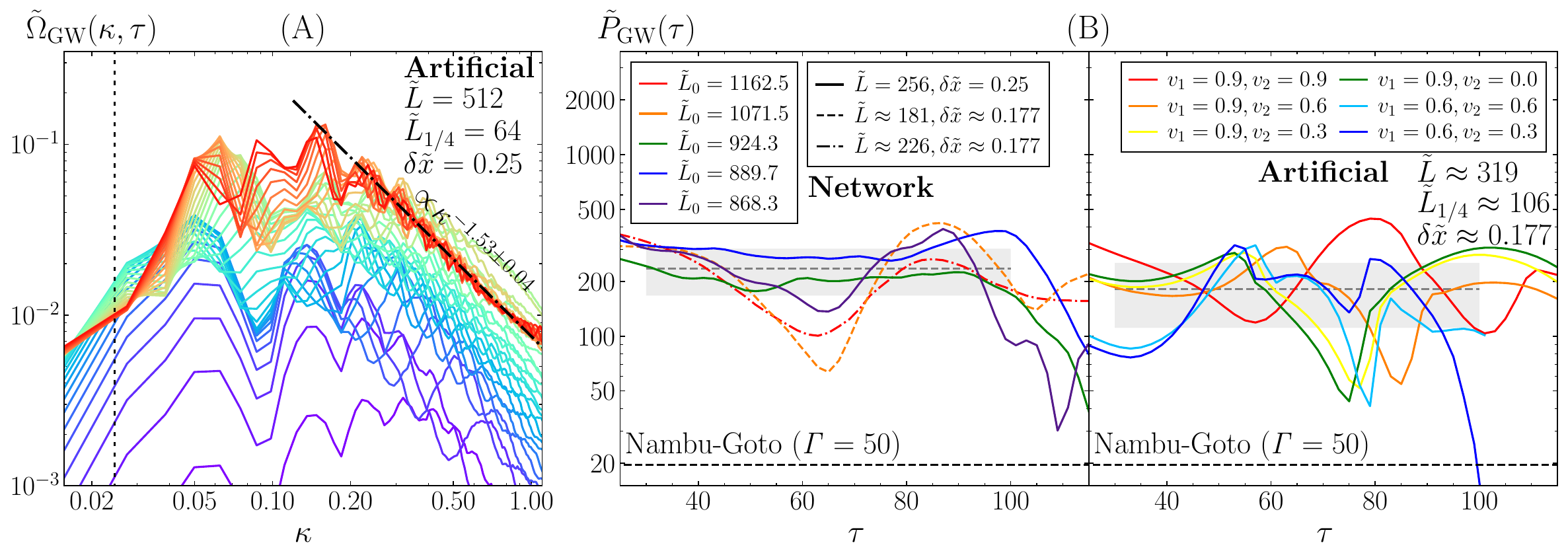}
     \caption{(A): Evolution of the GW energy density spectrum of an artificial loop 
generated with $v_1=0.6$, $v_2=0.7$ and $\sin \alpha=0.5$. 
Each colored-line corresponds to a different time, going from purple to red, with separations of two units of program time. The vertical dotted line indicates the scale of the initial string length and the dotted-dashed line is a fit to the high-frequency tail of the final-time spectrum. (B): Rolling-averaged GW power emitted by network loops of different length (left) and artificial loops  with distinct boost velocities (right), computed using Eq.~\ref{eq:averagedGWsPower}. For comparison the typical Nambu-Goto result (for $\mu=\pi v^2$ and $\mathit{\Gamma} =50$) is also shown.
}
     \label{fig:GWresults}
 \end{figure*}

We also study the effect of varying the IR coverage for artificial loops. Fixing $\tilde{L}_{1/4}=64$, $\delta\tilde{x}=0.25$,  $v_1=0.6$, $v_2=0.7$ and $\sin\alpha=0.5$, we vary the box size. The resulting GW spectra are shown in Fig.~\ref{fig:UVIRspectra}-(C), multiplied by a factor that accounts for lattice-size dependencies. While large discrepancies are observed for the smallest box, the spectra converge as the box size increases. GW emission is suppressed for scales larger than the initial loop length $\kappa < \kappa_0$ (black dotted line).  A zoomed-in version of the spectra from the largest box is shown in Fig.~\ref{fig:GWresults}-(A), where we note the presence of various peaks in the spectrum. Although the peak structure resembles the harmonic pattern expected in NG strings, peak frequencies here are not in harmonic proportions, and their absolute and relative location actually varies between early (blue) and late (red) times. 
We have fitted the high-frequency tail of the spectrum (dotted-dashed line), which approximately scales as $\propto k^{-3/2}$, representing a steeper fall than the standard Nambu-Goto (NG) tail, $\propto k^{-4/3}$, for cusp-dominated GW emission. 

We determine a ({\it rolling average}) measurement of the total GW power emitted by a loop as
\begin{eqnarray}\label{eq:averagedGWsPower}
P_{\rm GW}(t) \equiv 
\frac{L^3\rho_{\varphi}}{2T}\int^{t+T}_{t-T}\text{d}t'\int_0^{k_\text{cut}} \dot\Omega_\text{GW}(k,t')\,\text{d}\log k\,, 
\end{eqnarray}
which in terms of program variables can be written as $\tilde P_{\rm GW} = (P_{\rm GW}/v^2)\,/\,(v/m_\text{p})^2$. This is shown in Fig.~\ref{fig:GWresults}-(B) for $\tilde{T}\equiv T/\sqrt{\lambda} v=15$ and $\kappa_\text{cut}=1$. The latter is chosen to prevent inadvertently 
capturing the artificial UV-peak. 

The left panel of Fig.~\ref{fig:GWresults}-(B) corresponds to network loops of different $\tilde L_0$, simulated in lattices with varying $\tilde{L}$ and $\delta\tilde{x}$.  
In all cases the GW power emitted does not depend on $\tilde L_0$ and is roughly constant in time, with fluctuations that depend on the specific details of the dynamics of each loop. At late times  $\tilde P_{\rm GW}$ drops off as the loops finally disappear. Remarkably, there is no evidence of any systematic variation of the GW power emitted due to the shrinking of the loops. Between $\tau=30-100$, the average emission of all loops is $\tilde P_{\rm GW} = 240 \pm 80$ (grey dashed line and band). Although there is no a priori reason to expect this to be similar to the NG prediction, $P_\text{NG} = (\mathit{\Gamma}/8\pi)\mu^2/m_\text{p}^2$, it is still instructive to make the comparison. Using $\mu=\pi v^2$ and $\mathit{\Gamma} = 50$~\cite{Blanco-Pillado:2017oxo}, one gets $\tilde P_\text{NG} = \pi\mathit{\Gamma}/8 \approx 20$, 
roughly an order of magnitude smaller than our result (horizontal dashed line). 

Results for artificial loops with different boost velocities are presented in the right panel of Fig.~\ref{fig:GWresults}-(B). All simulations are performed using $\tilde{L}=450/\sqrt{2}\approx319$ and $\delta\tilde{x}=0.25/\sqrt{2}\approx0.177$, with initial string separation $\tilde{L}_{1/4}=\tilde{L}/3$, chosen to reduce IR effects. For all choices of $\lbrace v_1, v_2\rbrace$, 
the GW power emission is of a similar order as for network loops, again with an amplitude that does not show large variations as the loops shrink. 
Fluctuations are observed, which we believe are in part related to superposition of different GW fronts arising from the very peculiar symmetry of the squared configuration of the loops. Between $\tau=30-100$ the average emission is of the order $\tilde P_{\rm GW} = 190 \pm 80$ (grey dashed line and band), again an order of magnitude larger than the NG result. 

Comparing the GW and particle emission rates for both types of loops leads to
\begin{eqnarray}
{P_{\rm GW}\over P_{\varphi}} \approx \left\lbrace 
\begin{array}{ll}
\displaystyle{(240 \pm 80)\over (11.2 \pm 1.6)}\left({v\over m_\text{p}}\right)^2\,,&{\rm Network~loops,}\vspace*{1mm}\\
\displaystyle{(190 \pm 80)\over (5.2 \pm 2.5)}\left({v\over m_\text{p}}\right)^2\,,&{\rm Artificial~loops,}
\end{array}
\right.
\end{eqnarray}
where $P_\varphi$ for artificial loops is taken as an average over all studied families. 
As the string scale cannot be arbitrarily large, e.g.~CMB constraints require $v/m_\text{p} \lesssim 10^{-6}-10^{-3}$~\cite{Lopez-Eiguren:2017dmc,  Benabou:2023ghl}, we conclude that ${P_{\rm GW}/P_{\varphi}} \ll 1$, indicating that loop decay into GWs is completely sub-dominant compared to particle emission. This justifies a posteriori neglecting backreaction of GWs onto the loops.

 \section{Discussion and conclusions}\label{sec:conclusion}
 
 In this work, we study the decay of global string loops into scalar particles and GWs. We find 
the former totally dominates the decay for any acceptable vacuum expectation value of the strings, $v$, 
independently of their shape and initial conditions, 
with a universal suppression of GW to particle radiation as 
\begin{eqnarray}
{P_{\rm GW}\over P_{\varphi}}  \approx \mathcal{O}(10)\left({v\over m_\text{p}}\right)^2 \ll 1\,.
\end{eqnarray} 
Our lattice study shows that the above conclusion is robust for length-to-width ratios $80 \lesssim \tilde{L}_0/\tilde r_{\rm c} \lesssim 1700$,
with no indication that this would change for larger ratios.

Extrapolating our results to cosmological scales 
provides a new approach 
to calculate the GWB spectrum from a global string network. While current attempts are based on lattice simulations of entire networks~\cite{Figueroa:2012kw,Figueroa:2020lvo,Gorghetto:2021fsn}, or on a combination of field theory and NG ingredients~\cite{Chang:2019mza,Gouttenoire:2019kij,Chang:2021afa,Gouttenoire:2021jhk}, we suggest obtaining the GWB spectrum from a convolution of the loop number density at cosmological scales (see e.g.~\cite{Blanco-Pillado:2013qja,Klaer:2017qhr,Martins:2018dqg,Ringeval:2005kr,Auclair:2019zoz,Auclair:2021jud}) with our newly calibrated GW power emission. 
Crucially,
we have determined the GW emission of individual loops
without having to resolve 
the separation $d_\text{s}$ between strings in the network, and without losing resolution of the string core as field evolution progresses. Taking into account that the GW power emission we obtain is independent of the string length and scales as $\propto v^2(v/m_\text{p})^2$---and not proportional to a logarithmically growing tension as $\propto v^2(v/m_\text{p})^2\log^2(d_\text{s}/r_\text{c})$---together with the fact that global string loops are short lived, we anticipate that our results point to a suppression of the overall amplitude of the GWB from global cosmic string networks. We will present the details of this in a separate publication.\vspace{0.2cm}

\begin{center}
    {\bf Acknowledgments}
\end{center}
We are grateful to Tanmay Vachaspati and Geraldine Servant for useful conversations. JBB is supported by the Spanish MU grant FPU19/04326, from the European project H2020-MSCA-ITN-2019/860881-HIDDeN and from the staff exchange grant 101086085-ASYMMETRY. EJC is supported by a STFC Consolidated Grant No.~ST/T000732/1, and by a Leverhulme Research Fellowship RF-2021 312. DGF is supported by a Ram\'on y Cajal contract with Ref.~RYC-2017-23493 and by the  Generalitat Valenciana grant PROMETEO/2021/083. JL acknowledge support from Eusko Jaurlaritza IT1628-22 and by the PID2021-123703NB-C21 grant funded by MCIN/AEI/10.13039/501100011033/ and by ERDF; ``A way of making Europe”. We also acknowledge support from the grant PID2020-113644GB-I00. This work has been possible thanks to the computing infrastructure of Tirant and LluisVives clusters at the University of Valencia, FinisTerrae III at CESGA, and COSMA at the University of Durham.


\appendix

\section{Simulation parameters}\label{app:simulation}

This appendix summarizes the parameters used in the simulations. Table~\ref{tab:networkIC} refers to network loops, characterized by $\{\tilde L, \delta\tilde x, \tilde{\ell}_\text{str}\}$, while Table~\ref{tab:artificialIC} presents the initial conditions (upper table) and simulation parameters $\{\tilde{L}, \tilde{L}_{1/4}, \delta\tilde{x}\}$ (lower table) used for artificial loops. Note that in this case all sets of parameters are used for each family of initial conditions.
    
\begin{table*}[t!]
\begin{minipage}{\columnwidth}
\centering
\renewcommand{\arraystretch}{1.3}
\begin{tabular}{|>{\centering\arraybackslash}p{2.7cm}|>{\centering\arraybackslash}p{2.75cm}|>{\centering\arraybackslash}p{0.7cm}|>{\centering\arraybackslash}p{1cm}|}
\hline
 $\tilde{L}$ & $\delta\tilde{x}$ & $\tilde{\ell}_\text{str}$ & $\tilde{L}_0$   \\ \hline \hline 
 256 & 0.25 & 25 & 868.3 \\\hline
 256 & 0.25 & 20 & 924.3 \\\hline
 256 & 0.25 & 20 & 889.7 \\\hline
 $320/\sqrt{2}\approx226.3$ & $0.25/\sqrt{2}\approx0.177$ & 22 & 1162.48 \\\hline
 $256/\sqrt{2}\approx181.0$ & $0.25/\sqrt{2}\approx0.177$ & 15 & 1071.5 \\\hline
 $256/\sqrt{2}\approx181.0$ & $0.25/\sqrt{2}\approx0.177$ & 15 & 684.0 \\\hline
 $256/\sqrt{2}\approx181.0$ & $0.25/\sqrt{2}\approx0.177$ & 15 & 371.2 \\\hline
 144 & 0.25 & 12 & 557.0 \\\hline
 144 & 0.25 & 12 & 296.3 \\\hline
 144 & 0.1875 & 12 & 605.0 \\\hline
 144 & 0.1875 & 12 & 553.5 \\\hline
 144 & 0.125 & 12 & 633.6 \\\hline
 $192/\sqrt{2}\approx135.8$ & $0.25/\sqrt{2}\approx0.177$ & 15 & 728.6 \\\hline
 $192/\sqrt{2}\approx135.8$ & $0.25/\sqrt{2}\approx0.177$ & 15 & 699.8 \\\hline
 $192/\sqrt{2}\approx135.8$ & $0.25/\sqrt{2}\approx0.177$ & 15 & 582.2 \\\hline
 $192/\sqrt{2}\approx135.8$ & $0.25/\sqrt{2}\approx0.177$ & 15 & 462.4 \\\hline
 $192/\sqrt{2}\approx135.8$ & $0.25/\sqrt{2}\approx0.177$ & 15 & 405.2 \\\hline
 128 & 0.25 & 10 & 433.2 \\\hline
 $128/\sqrt{2}\approx90.5$ & $0.25/\sqrt{2}\approx0.177$ & 8 & 260.7 \\\hline
 $128/\sqrt{2}\approx90.5$ & $0.25/\sqrt{2}\approx0.177$ & 8 & 234.1 \\\hline
 $128/\sqrt{2}\approx90.5$ & $0.25/\sqrt{2}\approx0.177$ & 8 & 215.2 \\\hline
 $128/\sqrt{2}\approx90.5$ & $0.25/\sqrt{2}\approx0.177$ & 8 & 88.2 \\\hline
 $128/\sqrt{2}\approx90.5$ & $0.25/\sqrt{2}\approx0.177$ & 8 & 60.3 \\\hline
\end{tabular}
\caption{Summary of the simulation parameters used to study the decay of network loops, together with the initial length of the isolated loops estimated from the number of pierced plaquettes taking the Manhattan effect into account.}
\label{tab:networkIC}
\end{minipage}
\begin{minipage}[!t]{\columnwidth}
\centering
\renewcommand{\arraystretch}{1.3}
\begin{tabular}[t]{|>{\centering\arraybackslash}p{0.7cm}|>{\centering\arraybackslash}p{0.7cm}|>{\centering\arraybackslash}p{1cm}|}
\hline
  $v_1$ & $v_2$ & $\sin\alpha$   \\\hline \hline 
  0.9 & 0.9 & 0.4 \\\hline
  0.9 & 0.6 & 0.4 \\\hline
  0.9 & 0.3 & 0.4 \\\hline
  0.9 & 0.0 & 0.4 \\\hline
  0.6 & 0.6 & 0.4 \\\hline
  0.6 & 0.3 & 0.4 \\\hline
  0.3 & 0.3 & 0.4 \\\hline 
\end{tabular}\vspace{0.5cm}
\begin{tabular}[t]{|>{\centering\arraybackslash}p{2.7cm}|>{\centering\arraybackslash}p{2.75cm}|>{\centering\arraybackslash}p{2.7cm}|}
\hline
 $\tilde{L}$ & $\tilde{L}_{1/4}$ & $\delta\tilde{x}$  \\ \hline \hline 
 $50/\sqrt{2}\approx35.4$  & $25/\sqrt{2}\approx17.7$ & $0.25/\sqrt{2}\approx0.177$ \\\hline
 $100/\sqrt{2}\approx70.7$ & $50/\sqrt{2}\approx35.4$ & $0.25/\sqrt{2}\approx0.177$\\\hline
 $150/\sqrt{2}\approx106.1$ & $75/\sqrt{2}\approx53.0$ & $0.25/\sqrt{2}\approx0.177$\\\hline
 $200/\sqrt{2}\approx141.4$ & $100/\sqrt{2}\approx70.7$ & $0.25/\sqrt{2}\approx0.177$\\\hline
 $250/\sqrt{2}\approx176.8$ & $125/\sqrt{2}\approx88.4$ & $0.25/\sqrt{2}\approx0.177$\\\hline
 $300/\sqrt{2}\approx212.1$ & $150/\sqrt{2}\approx106.1$ & $0.25/\sqrt{2}\approx0.177$\\\hline
 $400/\sqrt{2}\approx282.8$ & $200/\sqrt{2}\approx141.4$ & $0.25/\sqrt{2}\approx0.177$\\\hline 
\end{tabular}
\caption{Summary of the initial conditions (top) and simulation parameters (bottom) used to study the decay of artificial loops. For each family of initial conditions, a simulation is performed using each $\{\tilde L, \tilde{L}_{1/4}, \delta\tilde{x}\}$ set. All simulations have $\sin\alpha=0.4$.}
\label{tab:artificialIC}\vfill
\end{minipage}
\end{table*}

\bibliography{automatic,manual}

\end{document}